\documentclass[%
 aip,amsmath,amssymb,
 reprint,%
]{revtex4-1}

\usepackage{graphicx}
\usepackage{dcolumn}
\usepackage{bm}

\usepackage[utf8]{inputenc}
\usepackage[T1]{fontenc}
\usepackage{mathptmx}
\usepackage{etoolbox}

\usepackage{hyperref}
\hypersetup{
    colorlinks=true,
    allcolors=black,
    urlcolor=cyan,
    linkcolor=blue,
    filecolor=magenta,
    pdftitle={Practice Makes Better},
    pdfpagemode=UseOutlines,
}

\makeatletter
\def\@email#1#2{%
 \endgroup
 \patchcmd{\titleblock@produce}
  {\frontmatter@RRAPformat}
  {\frontmatter@RRAPformat{\produce@RRAP{*#1\href{mailto:#2}{#2}}}\frontmatter@RRAPformat}
  {}{}
}%
\makeatother

\usepackage[export]{adjustbox} 

\usepackage{color} 
\definecolor{bleudefrance}{rgb}{0.19, 0.55, 0.91}
\definecolor{AmericanRed}{rgb}{0.698, 0.133, 0.204}
\definecolor{AmericanBlue}{rgb}{0.0391, 0.1914, 0.3789}
\definecolor{mustard}{rgb}{.808,.702,.00392}
\definecolor{ketchup}{rgb}{0.781, 0.160, 0.1328}
\definecolor{beeswax}{rgb}{.9137, .6235, .2471} 
\definecolor{grey}{rgb}{.5,.5,.5}
\definecolor{orange}{rgb}{1,.357,0}

\newcommand{\Nqtot}{\ensuremath{N_{\rm Q,tot}}}
\newcommand{\Nsess}{\ensuremath{N_{\rm sess}}}
\newcommand{\Nqbar}{\ensuremath{N_{\rm Q, mean}}}

\newcommand*{\wm}[1]{\textcolor{AmericanRed}{\textsf{WKB: #1}}}

\begin{document}

\preprint{AJP}

\title[Practice Makes Better]{Practice Makes Better: 
Quantifying Grade Benefits of Study}

\author{\href{https://orcid.org/
0000-0003-4811-7913}{W. K. Black}}
  \email{wkblack@umich.edu}
  \affiliation{Departments of Physics and Astronomy \& Astrophysics,  University of Michigan, Ann Arbor}

\author{Rebecca L. Matz}
\author{Mark Mills}
  \affiliation{Center for Academic Innovation, University of Michigan, Ann Arbor}

\author{A. E. Evrard}
  \affiliation{Departments of Physics and Astronomy \& Astrophysics,  University of Michigan, Ann Arbor}
  \affiliation{Program in Computing for the Arts and Sciences,  University of Michigan, Ann Arbor}

\date{\today}

\begin{abstract}
Problem Roulette (PR), an online study service at the University of Michigan, offers points-free formative practice to students preparing for examinations in introductory STEM courses. Using four years of PR data involving millions of problem attempts by thousands of students, we quantify benefits of increased practice study volume in introductory physics. 
After conditioning mean final grade on standardized (ACT/SAT) math test score, we analyze deviations based on student study volume.  We find a strong effect; mean course grade rises quadratically with the logarithm of the total number of PR questions encountered over the term (\Nqtot), with an overall gain of $0.77 \pm 0.12$ grade points between $1 < \Nqtot < 1000$.  The gains are persistent across the range of math test score represented in our sample.
While \Nqtot\ surely correlates with other study habits, the benefits of increased study in general still hold. 
A model for final grade using test score and study volume largely accounts for demographic stratification, including by sex, parental education level, number of parents at home, nationality / underrepresented minority status, and regional income level, with two significant exceptions: students whose parents did not earn a college degree, who earn $-0.27 \pm 0.04$ grade points ($6.1\sigma$) below expectations and underrepresented minority students at $-0.14 \pm 0.04$ points ($3.6\sigma$). Residual scatter in final grade remains comparable to the maximal study gains, implying that the model is far from deterministic: individual variation trumps mean trends. 
Our findings can help motivate students to study more and help teachers to identify which types of students may especially need such encouragement. 
\end{abstract}

\maketitle

\section{Introduction} 

Physics is a notoriously difficult subject in the eyes of many undergraduates.\cite{Wong2022} 
Introductory physics courses are typically among the most-failed courses on college campuses.  According to student evaluations of teaching at our university\footnote{Made available to campus members by the Atlas service, \url{http:atlas.ai.umich.edu}.} the workload of the first-semester physics course is perceived as considerably heavier than the workload of introductory courses in general chemistry or statistics. In a study of past examination problems for these three subjects\cite{Weaverdyck+20}, we found that introductory physics questions are both \textsl{more complex} (take a longer time to solve on average) and \textsl{harder} (have a lower average correct response rate) than questions in chemistry and statistics. Asked by students how to succeed in physics, instructors often recommend practicing more problems, among other strategies. We seek here to measure how such practice study affects final grade.

Benefits of practice study are well-documented, though some forms of study benefit students more than others.
In 2013, \citet{Dunlosky+13} performed a meta-analysis on hundreds of studies on learning, measuring the utility of various study methods. Of the ten main study methods considered, 
they found that \textbf{practice testing} (``self-testing or taking practice tests over to-be-learned material'') and \textbf{distributed practice} (``implementing a schedule of practice that spreads out study activities over time'') were of the highest utility. Both methods benefited learners of different ages and abilities and were shown to boost students' performance across many different kinds of tasks and contexts. (In contrast, study methods such as summarizing, highlighting, and rereading were less effective.) 
A more recent meta-analysis\cite{Yang+21} of several hundred studies finds significant gains from testing (a term they use to include quizzing), measuring boosts in student attainment that benefit all kinds of students in similar manner. 
Such findings help motivate use of a study tool at the University of Michigan known as Problem Roulette.\cite{Evrard2015PR} 

\begin{figure}\centering
  \includegraphics [width=.75\linewidth, frame] {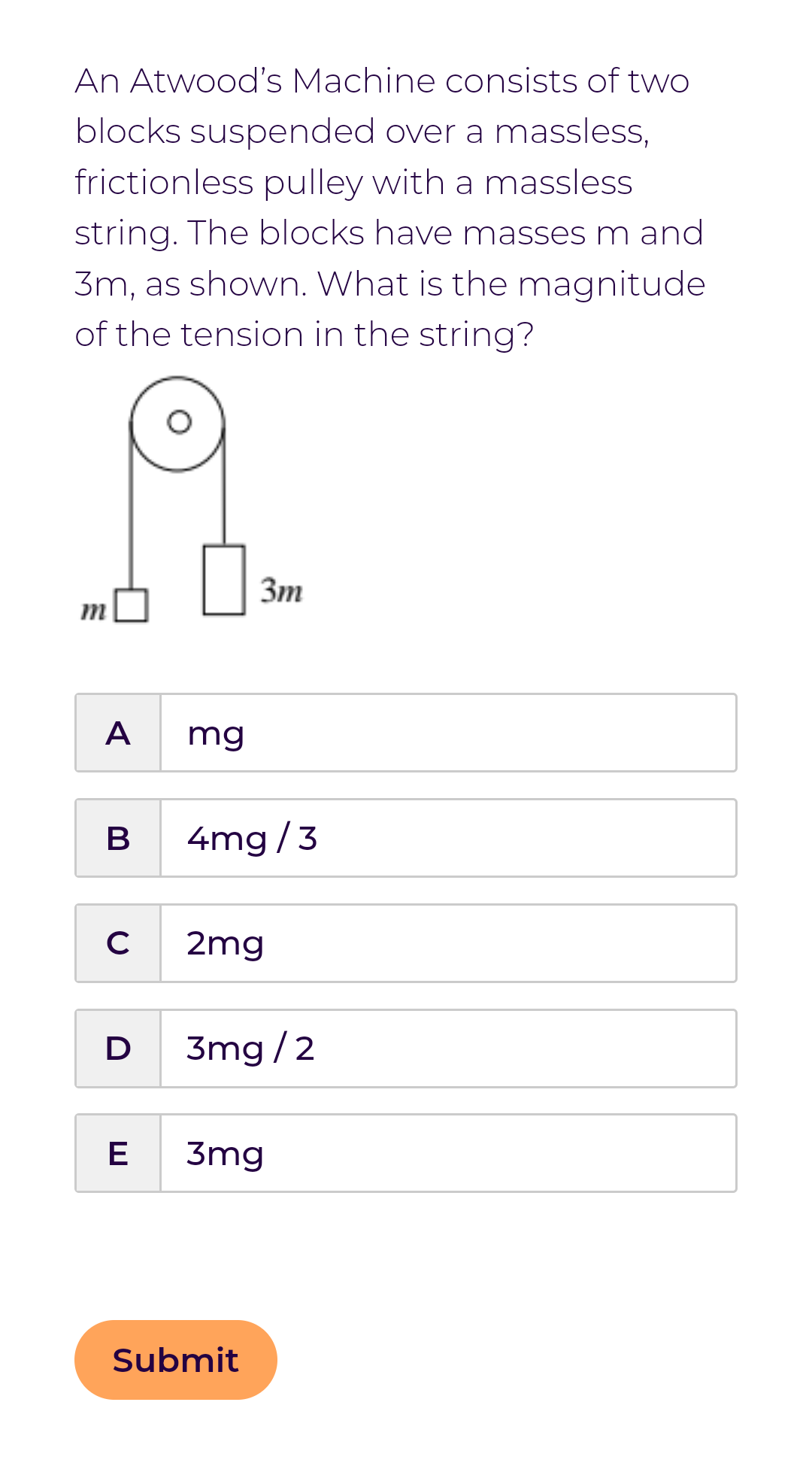}
  \caption{
    Example Problem Roulette question as viewed on a mobile device for the course PHYSICS 140. 
  }
  \label{fig:PR_Q}
\end{figure}

\href{https://problemroulette.ai.umich.edu/about}{Problem Roulette} (PR) is an optional, points-free study service at the University of Michigan (UM) that provides students open access to a large library of locally-authored, topically-organized  problems in multiple subject areas. Most of the content consists of multiple-choice questions used in past examinations in introductory science, technology, engineering, and mathematics (STEM) courses. 
``Roulette'' refers to both its random selection of questions 
(an example is shown in Figure~\ref{fig:PR_Q}) as well as its reflection of high-risk assessments. Students can optionally activate settings which simulate timed tests, mirroring the difficulty and stress of actual exams; see Appendix~\ref{apx:PR} for more details on study modes and instructor options. 
PR's equality of accessibility to UM students sidesteps paywalls of online repositories and bypasses access limits to large banks of past exams held by exclusive student groups such as fraternities, sororities, and honor societies. 

Our group previously examined study behaviors and grade benefits of practice based on PR usage data from 2013 to 2017 in three subject areas. \citet{Weaverdyck+20} measured study gains using a median split of high vs low study, finding in the high study group a moderate benefit of $\sim 0.15$ grade points in chemistry and statistics and a weaker benefit in physics. 
They also found that female students worked $\sim 25\%$ more problems on average than male students in all three subjects; they additionally received a slightly higher grade benefit than male students from higher volumes of PR practice.  

To quantify these study benefits they used a measure called ``grade anomaly'', comparing each student's grade points earned (GPE) in a particular course relative to their end-of-term cumulative GPA computed from all \textsl{other}  classes (termed GPAO).\cite{Huberth+15,Koester+16} ``Grade anomaly'' is then simply ${\rm GPE} - {\rm GPAO}$. 
While measures such as GPAO are commonly used to control for student performance, ``grade anomaly'' masks a large portion of grade benefit from practice study due to its correlation with study habits (see \S\ref{sec:GPAO}), so subtracting off GPAO from GPE subsumes a significant portion of otherwise measureable study gains. It is therefore not optimal for our current study, which seeks to quantify gains of study. 



In this paper, after modeling GPE as a function of standardized math exam score (concordant ACT and SAT math subscores), we quantify the extent to which deviations from the mean trend correspond to increased PR study volume. 
We then consider five student demographic characteristics: sex, parental education level, single parent status, nationality and underrepresented minority (URM) status, and high school zip code median income (a proxy for estimating family income). 
Grade differences within each of these subgroups have been noted; linear regression and studies of linear correlations have shown that these differences in grades are ameliorated to varying degrees on accounting for test scores, personality, study habits, and other factors.\cite{Walpole_2003,Aluja_Blanch_2004,Pascarella+04,Delaney+11,Richardson+12,McLanahan+13,Matz+17,Simmons+20} 
In contrast, in this paper we use the non-linear tool KLLR to investigate to what extent differences between these demographic sub-groups are explained or exacerbated by accounting for test scores and study volume. 

We describe our methods and data in \S\ref{sec:methods} before presenting our key findings in \S\ref{sec:results}.  
  That section begins with grade gains from practice conditioned on ACT/SAT math score (\S\ref{sec:mu_deviation_study}) followed by an exploration of demographic differences (\S\ref{sec:mu_deviation_demo}) and a comparison to grade gains conditioned on GPAO (\S\ref{sec:GPAO}).  
We offer reflection in \S\ref{sec:discussion} and conclude with succinct recommendations for teachers and students in \S\ref{sec:summary}.


\section{Methods} \label{sec:methods}

In this section, we describe the flexible population modeling method known as kernel-localized linear regression (KLLR), introduce the data and scope of our study, demographic groupings, and introduce the math test score measure ($T$) used to model mean GPE as a base condition.

\subsection{Measuring trends with KLLR} \label{sec:KLLR}
While any differentiable function is linear at small enough scales, life tends to be non-linear. As one considers increasingly wider domains, behavior of any variable moves from a simple mean to a linear trend to a quadratic one and so forth (as is the nature of Taylor series). Our data in student math scores and study habits cover wide enough ranges that simple averages and even linear fits become statistically disfavored. 

Furthermore, there are reasons to expect significant non-linearity in the trends we consider in this paper. One might expect grade points earned (GPE) as a function of math test score to have a sigmoid relation (if test scores ranged high enough), as there's a definite floor and ceiling to GPE (and a practical floor of grades given or accepted, as only some grades are passing). 
One might expect complex behavior in the space of grade benefit versus study volume in a given term, showing steady increases until a student begins to overstudy (eventually even foregoing sleeping and eating to increase study volume), followed by a decrease in performance. These expected non-linearities motivated us to use a statistical method beyond simple averages, linear trends, or even polynomial fitting---a method capable of measuring local, non-linear trends in a relatively agnostic fashion.

To examine mean relationships of grades and grade benefits as smooth functions of secondary variables, we employ Kernel-Localized Linear Regression\cite{Farahi+18,Farahi+22} (KLLR), a method that determines parameters of a locally linear fit (mean, slope, and variance) within a sliding Gaussian window. This approach to population modeling, originally developed in the 1970s,\cite{cleveland1979robust, cleveland1996smoothing, takezawa2005introduction} allows for more nuanced analysis than polynomial fitting as it does not enforce a particular global behavior.

A single parameter is required for the method, the width of the Gaussian filter. As the Gaussian window slides across the domain, KLLR produces continuous and smooth fits to data. The kernel widths are chosen to be roughly a fifth of their range, 
\begin{equation} \label{eqn:KLLR_kw} 
  \sigma_{\rm KLLR} = \frac15 (q_{99\%} - q_{1\%})
\end{equation}
where $q_{n\%}$ is the $n$th quantile of the logged data. 
For our data, this kernel size minimizes noise in the fit while maximizing allowed internal variation, lying on the threshold of non-monotonic behavior.

\subsection{Courses investigated} \label{sec:courses}

Problem Roulette currently supports sixteen courses with over ten thousand unique questions available in aggregate. Our primary focus here is on the introductory, calculus-based physics sequence for scientists and engineers:  
  PHYSICS 140 (General Physics I: Mechanics) and its continuation, 
  PHYSICS 240 (General Physics II: Electricity and Magnetism). 
These courses are primarily ($\sim 80\%$) taken by engineering students. 
\footnote{UM also offers a similar sequence designed for life sciences students as well as an honors sequence for physics majors. The former sequence was rebuilt during the study period and so is not yet supported by PR.  The latter sequence has a much smaller enrollment which limits its statistical power.}

While our focus is on study in physics, in this section only we offer some context by including basic PR usage and mean grade behavior in three other PR-supported STEM courses: General Chemistry: Macroscopic Investigations and Reaction Principles (CHEM 130), Elementary Programming Concepts (EECS 183), and  Introduction to Statistics and Data Analysis (STATS 250). 

\begin{table}[!ht]
\centering
  \caption{
    Usage statistics for selected PR-supported courses over a seven-semester period (Winter 2018 to Winter 2021). We focus on PHYSICS 140 and 240 here but include three other courses for context. Statistics reflect the $\sim 93\%$ of students with ACT or SAT scores available. 
  }
  \begin{ruledtabular}
    \begin{tabular}{lccccc}
      Course & $N_{\rm Q, course}$\footnote{Number of unique questions available.} & $N_{\rm students}$\footnote{Number of students enrolled during the study period.} & $f_{\rm use}$\footnote{Fraction of students encountering at least one question on PR.} & $\sum N_{\rm Q, tot}$\footnote{Total number of questions completed per term by all students.} & $\sum \Nsess$\footnote{Total number of sessions completed per term by all students.} \\
      \hline
      PHYSICS 140 & 854 & 3856 & 82.5\% & 348208 & 27471 \\ 
      PHYSICS 240 & 931 & 2772 & 60.8\% & 131868 & 10312  \\ 
      \hline
      CHEM 130 & 685 & 5786 & 90.0\% & 1036986 & 59358 \\ 
      EECS 183 & 953 & 3803 & 71.3\% & 246239 & 13371  \\ 
      STATS 250 & 526 & 9435 & 74.9\% & 614700 & 38242 \\
    \end{tabular}
  \end{ruledtabular}
  \label{tab:courses}
\end{table}

Table~\ref{tab:courses} shows PR usage statistics for the seven-semester study period beginning Winter 2018 and ending Winter 2021 (inclusive). Spring and Summer terms during this period are also included, but these have much smaller enrollments than the traditional Fall/Winter terms. The student count includes only those with either an ACT or SAT scores recorded, representing $\sim 93\%$ of the full student enrollment.  
%


Because it is an optional service that is not employed for summative course assessments, not all students use PR.  The usage fractions are generally high, with values ranging from 60\% in PHYSICS 240 to 90\% in CHEM 130.  Note that the student numbers are unique only within each class; many students enroll in several of these courses over the course of their careers.  The intensity of study is highest in CHEM 130, with an average of 200 questions attempted per term by PR-using students. 
By this metric, study volume in PHYSICS courses is considerably lower, with a mean value of 110 and 78 questions per term in 140 and 240, respectively. 
Median study volumes for physics courses were also smaller by more than a factor of two compared to CHEM~130.

The study was determined to be exempt from ongoing review by our Institutional Review Board (HUM00158291).

\subsection{Study volume indicators} \label{sec:study_indicators}
\begin{figure}[!t]
\centering
  \includegraphics[width=\linewidth]{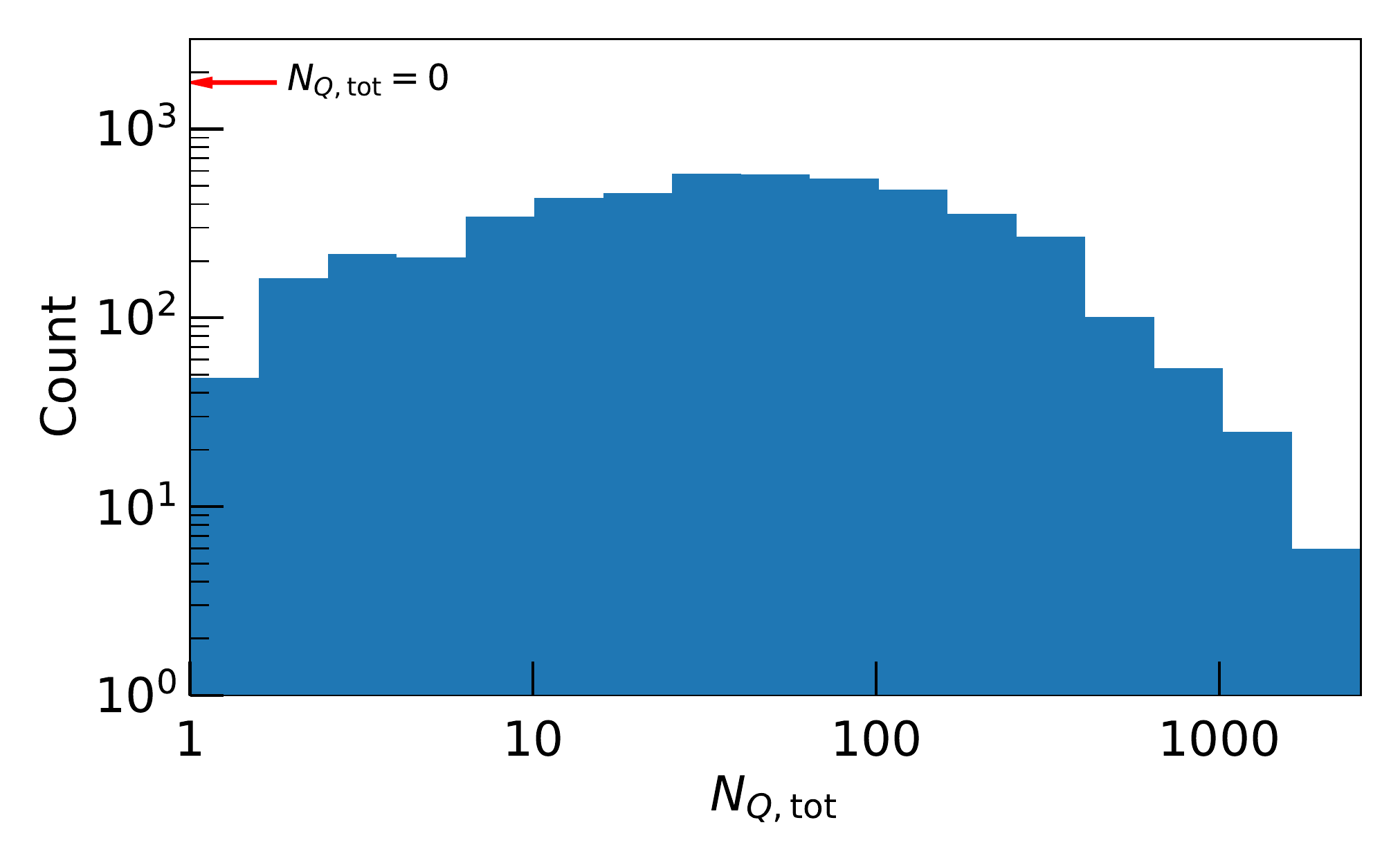}
  \caption{
    Histogram of total unique questions encountered per session, \Nqtot, for Physics 140 and 240 combined. Red arrow indicates the number of students who didn't use PR. 
  }
  \label{fig:hist_NQtot}
\end{figure}

We use several indicators of study volume in this paper, each calculated on a per term basis and summarized in Table~\ref{tab:study_indicators}.  
The primary measure we use is the total number of unique questions encountered per session, summed over all sessions on Problem Roulette. This measure, $N_{\rm Q, tot}$, is given for each student individually over the full academic term.  
As shown in Figure~\ref{fig:hist_NQtot}, the distribution of \Nqtot\ is close to log-normal (as are the distributions of \Nsess\ and \Nqbar).
For the combined sample of PR users in both physics courses, the log-mean value of \Nqtot\ is $\exp{\langle \ln \Nqtot \rangle } \simeq 37.5$, with a factor of 4.3 standard deviation. This value includes skipped questions, which represent roughly 20\% of the total of all questions encountered.

\begin{table}[!ht]\centering
  \caption{
    Study volume indicators for PHYSICS classes described in the text. Values are the 50\% (median), 75\%, and 95\% quantiles of each quantity for a sample of students with non-zero PR study (see Table~\ref{tab:courses} for usage fractions) and available ACT/SAT math scores (93\% of total enrollment). 
  }
  \begin{ruledtabular}
    \begin{tabular}{l ccc ccc}
      Indicator & \multicolumn{3}{c}{PHYSICS 140} & \multicolumn{3}{c}{PHYSICS 240} \\ 
     &  50\% & 75\% & 95\% & 50\% & 75\% & 95\% \\ 
      \hline
    $\Nqtot$ & 50 & 103 & 389 & 25 & 36 & 296 \\ 
    $\Nsess$ & 5 & 9 & 28 & 3 & 4 & 21 \\ 
    $\Nqbar$ & 9.5 & 12.4 & 26 & 8.3 & 10 & 24 \\ 
    \end{tabular}
  \end{ruledtabular}
  \label{tab:study_indicators}
\end{table}

We also examine two additional measures of practice: i) the number of PR sessions, $\Nsess$, where a new session is triggered when a student logs on and responds to at least one question, either for the first time or after a period of inactivity since responding to the last question; and ii) the mean number of questions encountered per session, $\Nqbar$.  

Over the study period, different approaches to midterm and final examinations were taken by instructors.  In some terms three evening exams, spaced by roughly one month, were held, each with typically 20 questions.  In other terms, biweekly 10-question ``quizzes'' were held. In some semesters these were held during class meetings while in others (particularly during the COVID-19 shutdown period) they were held in the evening.  Final examinations were cumulative and typically consisted of 25 questions. 
All questions were multiple-choice (typically five possible responses), authored by the instructors for each exam.  While some degree of similarity between exam questions and PR questions is likely, direct re-use of PR questions is strongly disfavored by the departmental culture of testing students with new questions.\footnote{Further evidence against PR practice as ``teaching to the test'' comes from exam scores, which average $\sim 65\%$ with a dispersion of $15\%$.}
All examinations were proctored and timed, and across the study period student scores on these assessments contributed half or slightly more to their final letter grade. Overall weighted student scores map to letter grades in a manner defined at the beginning of the semester; there is no ``curve'' in the traditional sense of using ranked ordering to define grade boundaries.  

With that context, the $\Nsess$ values in Table~\ref{tab:study_indicators} reflect the fact that PR-engaged students typically use the service once per examination, but with quite a large spread.  Combining both courses, we find log-mean number of sessions $\exp{\langle \ln \Nsess \rangle } \simeq 4.4$, with nearly a factor of three dispersion.  Five percent of students had 25 or more sessions over the term, a frequency of roughly two sessions per week. 
Sessions typically average about 30 minutes in duration, during which students attempt roughly eight problems.  Again, the dispersion is large (factor $\sim 3$), and five percent of students average 25 problems per session.  


Our analysis using PR-based statistics is an incomplete proxy for total study effort by students.  
Students do use other methods of study, including doing homework problems from their online textbook service,  
using tutors, or by going to on-campus sites such as the Physics Help Room and Science Learning Center. 
While inherently only a partial picture of student study effort, the volume of PR activity is substantial enough to offer useful insight, as we show below (see Fig.~\ref{fig:deviation_140240}). 
Furthermore, because these study indicators correlate with study habits in general, they may even be seen as proxies for studying of any kind, pointing towards effects beyond PR activity alone.

\subsection{Demographic groups} \label{sec:demo_groups}

We examine several demographic distinctions, delineated in Table~\ref{tab:demographics}. Table~\ref{tab:demo_pcts} gives the fractions for each group among the ACT/SAT student populations in the two physics courses of interest.

\begin{table*}\centering
  \caption{
    Demographic categories and sub-groups considered in this analysis. 
  }
  \begin{ruledtabular}
    \begin{tabular}{lll}
      Category & Sub-group & Description \\
      \hline 
      Sex & Male, Female & Sex as catalogued in LARC \\
      Parental education & $\leq$HS; A,B; M,D & 
      Highest degree earned: high school degree or less; associate's or bachelor's; master's or doctorate \\ 
      Parents at home & Both, Single & Whether the student comes from a single-parent household \\ 
      N/URM\footnote{Nationality/URM status} & URM, NUR, ITL & ITL if student is international; else URM=underrepresented minority, NUR=non-URM \\ 
      Income bin & $1^{\rm st}$, $2^{\rm nd}$, $3^{\rm rd}$, $4^{\rm th}$ & Median income of high school ZIP in near-quartile groups, divided by values of \$\{50, 75, 100\}k \\ 
    \end{tabular}
  \end{ruledtabular}
  \label{tab:demographics}
\end{table*}

\begin{table*}\small\sf\centering
  \caption{
    Fraction of course participants in each demographic subgroup. 
  }
  \begin{ruledtabular}
    \begin{tabular}{c|cc|ccc|cc|ccc|cccc}
    Course & Male & Female & $\leq$HS & A,B & M,D & Single & Both & URM & NUR & ITL & $1^{\rm st}$ & $2^{\rm nd}$ & $3^{\rm rd}$ & $4^{\rm th}$ \\
    \hline
    PHYSICS 140 & 64\% & 36\% & 10\% & 29\% & 60\% & 15\% & 85\% & 15\% & 80\% & 5\% & 22\% & 31\% & 18\% & 22\% \\
    PHYSICS 240 & 74\% & 26\% & 9\% & 27\% & 62\% & 13\% & 87\% & 12\% & 83\% & 5\% & 22\% & 32\% & 19\% & 21\% \\
    \end{tabular}   
  \end{ruledtabular}
  \label{tab:demo_pcts}
\end{table*}

The University of Michigan's definition of underrepresented racial/ethnic minority (URM) is tied to nationality and requires some exposition. 
As defined in the Learning Analytics Data Architecture (LARC)\cite{Lonn+19} \href{https://drive.google.com/file/d/0ByMPshcLeA_5dnc3dXhPR1FvZTA/view?usp=sharing&resourcekey=0-E-n-UoJBTvGYmMvtSKb4yg}{Data Dictionary}, students are considered international (ITL) if they are neither U.S. citizens nor U.S. permanent residents. 
The international student population is roughly two-thirds Asian, with the plurality of students from China ($46\%$) and another $5\%$ from the Republic of Korea. The rest of international students come to UM from over 100 countries, each individually accounting for $<4\%$ of the remainder. 
Of the remaining domestic student population, students are considered underrepresented if they self-identify as i) Black or African American, ii) Hispanic, iii) Native American, or iv) Native Hawaiian or Other Pacific Islander. Otherwise, they are considered non-underrepresented minorities (NUR). 
In the terms considered here, the URM population was chiefly composed of 
  Hispanic students ($42\%$), 
  students identifying as more than one ethnicity or race ($30\%$), and 
  Black students ($27\%$).

\subsection{ACT/SAT math score as a control condition} \label{sec:T}

The standard deviation of grades achieved in these courses is large, roughly $0.8$ on the standard $4.0$ scale.\footnote{Our university uses a standard A--E/F letter grade scale that maps to a numerical point scale of 4.0 (A) to 0 (E/F); B=3.0, C=2.0, D=1.0. UM also supports $\pm$ gradations that count for 0.3 grade point deviations, so e.g. a B+ is 3.3 grade points and a B- is 2.7 grade points.}  
A particular student's grade is influenced by a host of factors; we do not seek to (nor could we) capture all these factors simultaneously. 
Instead, we opt to employ a proxy for mathematical proclivity using a student's score on the ACT or SAT mathematics test. Recent work\cite{Simmons+20} on a large sample of introductory physics students identifies ACT math score as a strong predictor of final grade.  
Of the 7417 students enrolled in this introductory physics sequence during the study period, only 498 have neither ACT nor SAT scores recorded, meaning that 93\% of the total student sample have pre-college math test scores available.

%


In our analysis, we map ACT to SAT test scores using a 2018 concordance table from \href{https://downloads.compassprep.com/sat-act-score-concordance-2018.pdf}{Compass Prep}.\cite{CollegeBoard18}
Though fractions of students with SAT and ACT scores is similar, between PHYSICS 140 and 240, $3\%$ more students have SAT than ACT scores. We thus minimize imputation by converting scores from ACT to SAT. (Imputing in the opposite direction has negligible effects on results.) In cases where both scores were present, their average was taken. In cases where the same test was submitted multiple times, the highest score was used.  Noting that the SAT minimum component score is 200 and maximum 800, we map SAT math scores 
to the unit interval by defining 
\begin{equation}
    T_i = \frac{{\rm SAT}_i - 200}{800 - 200},
\end{equation}
where ${\rm SAT}_i$ is the actual, imputed, or averaged SAT math score for the $i^{\rm th}$ student. 

 
The distribution of $T$ leans heavily toward high values in our datasets, especially so in the physics courses. 
The two physics courses considered in our study have similar distributions of $T$, with medians near $0.92$, 10\% \& 90\% quantile values near $0.80$ \& $1.0$, respectively, and $\sim 0.44$ as their minimum value. A recent study on grade inflation \cite{Evrard_Schulz_Hayward_2021} at our university shows regular growth in ACT and SAT scores; rescaling SAT and ACT scores into the unit interval to match $T$, we see annual gains of $0.006$ in scores for both tests.  While this increase reflects 
a heightened selectivity of the institution, it is an order of magnitude smaller than the scatter in $T$ and therefore has relatively little effect over our study period.


While differences in mean scores have been noted for several demographic sub-groups historically as well as in our sample (see Figure~\ref{fig:mu_gpe_demo}), we find that intrinsic scatter within each sub-group far exceeds the distance between group mean scores. 
There is no definitive consensus among researchers as to the \emph{cause} of these differences in mean scores,\citep{Jensen_1980,Drasgow_1987,Herrnstein+1994,Jensen98,Brown+99,Frey+04,Koenig+08,Coyle+08,Dorans_2010,Soares+15,Geiser_2020} particularly whether this is a bias in measurement (i.e., whether two individuals with equal capacity but different demographic membership have significantly different scores) or a purely relational bias (i.e., the tests properly measure capacity, but differences in mean group scores are caused by some secondary variable). 
A commensurate summary of the conflicting interpretations of demographic differences in scores is beyond the scope of this paper. 
Because of the significant predictive power of ACT and SAT tests in estimating college grades, we find utility in using the measure $T$ as a baseline, and relegate investigation of its interpretation to future research.

\subsection{Modelling mean GPE versus test scores} \label{sec:mu_fit}

In this section, we describe how we condition grades on math test scores. To condition one variable on another is an attempt to remove its effects, in order to discern further trends. Here we subtract out mean grade points earned (GPE) at a given math test score $T$ in order to mitigate the strongly confounding factor of differences in math skills on expected grade. For example, if two students of vastly differing math skills both don't study, they are unlikely to earn identical grades. By conditioning on $T$, we subtract out the mean expected grade, allowing us then to measure how divergence from those expectations corresponds with study habits. 

\begin{figure}[!ht]\centering
  \includegraphics [width=\linewidth] {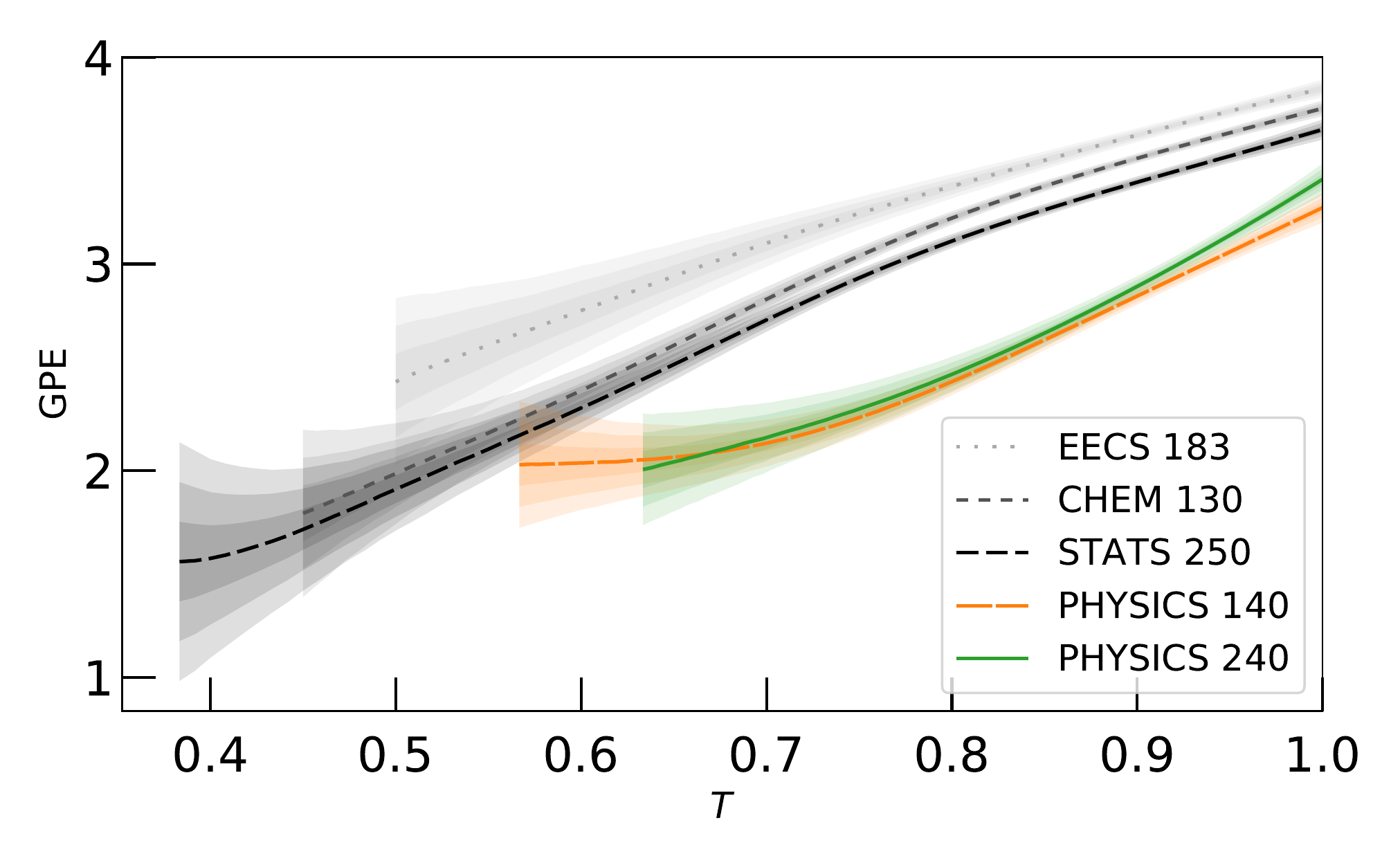}
  \caption{
    KLLR fits to mean grade points earned (GPE) as a function of normalized ACT/SAT math score ($T$), shown for multiple STEM courses (see legend). 
    Shaded regions show $\pm 3 \sigma$ uncertainties on the fit. 
    The left-side $T$ limits displayed are set by a limit of ten students. 
  }
  \label{fig:mu_gpe}
\end{figure}


We use KLLR to fit mean grade $\mu_{\rm GPE}$ as a function of $T$, allowing us to detect any potential non-linearities in trends. Using all student $T$ values, equation~\eqref{eqn:KLLR_kw} yields $\sigma_{\rm KLLR} = 0.09$, which value we use for all course KLLR fits, shown in Figure~\ref{fig:mu_gpe}. Though all five courses considered here are STEM courses, we find stark differences in the mean relationships between physics (concave up) versus non-physics courses (concave down). While students with higher math scores tend to score higher in all their courses, even students with above-average math scores of $T = 0.75$ tend to score lower grades in physics relative to the other STEM courses considered here.

For each course, we find that mean grade point earned as a function of $T$ is well-fit for the majority of students by a quadratic:
\begin{equation}\label{eqn:mu_gpe}
  \mu_{\rm GPE}(T) = a_0 + a_1 (T-T_0) + a_2 (T-T_0)^2
\end{equation}
(using pivot $T_0=0.9$, the physics median $T$ value). Parameter fits for each course are listed in Table~\ref{tab:mu_fits_new}. 

\begin{table}\small\sf\centering
  \caption{
    Fit constants quadratic functions $\mu_{\rm GPE}(T)$ (equation~\eqref{eqn:mu_gpe}) for each course, giving the mean trend of GPE (course grade points earned) as a function of $T$ (concordant ACT / SAT math subscores), determined using all students with ACT or SAT scores recorded. Given for each class individually as well as the merged set of PHYSICS 140 + 240. 
  }
  \begin{ruledtabular}\begin{tabular}{llll}
      Course & $a_0$ & $a_1$ & $a_2$ \\ 
      \hline
      PHYSICS 140 & $2.82 \pm 0.02$ & $4.3 \pm 0.2$ & $+5.3 \pm 1.2$ \\
      PHYSICS 240 & $2.82 \pm 0.02$ & $5.0 \pm 0.2$ & $+8.9 \pm 1.7$  \\
      MERGED & $2.82 \pm 0.01$ & $4.6 \pm 0.1$ & $+6.7 \pm 1.0$  \\
      \hline 
      CHEM 130  & $3.53 \pm 0.01$ & $2.6 \pm 0.1$ & $-3.8 \pm 0.5$  \\
      EECS 183  & $3.63 \pm 0.01$ & $2.3 \pm 0.2$ & $-1.8 \pm 0.7$ \\
      STATS 250 & $3.39 \pm 0.01$ & $2.8 \pm 0.1$ & $-3.4 \pm 0.5$  \\
  \end{tabular}\end{ruledtabular}
  \label{tab:mu_fits_new}
\end{table}

We note that when we measured fit parameters using only those students who did not use PR for study, we found good consistency in slopes and curvatures (hinting that study trends don't correlate strongly with $T$; see lower panel of Figure~\ref{fig:mu_gpe_demo} and Figure~\ref{fig:N_sess__vs__GPAO__by__T}), but intercepts $a_0$ were lower by $0.10 \pm 0.04$ grade points. This is not surprising when one considers that, overall, those who didn't study on PR in PHYSICS 140 had lower grades by $0.14 \pm 0.04$ points and in PHYSICS 240 by $0.10 \pm 0.03$ points. This vertical shift is precisely what this study seeks to quantify; by subtracting out $\mu_{\rm GPE}(T)$, one can then measure these divergences from expectations. These significant shifts hint already at the benefits of study. 

For the two physics courses, we found no significant differences in slope nor curvature for any demographic distinction. Though we found several significant differences in vertical shift between sub-groups, as explained above, these shifts do not detract from our analysis, as they are precisely what we set out to investigate by subtracting out a $\mu_{\rm GPE}(T)$ baseline. See \S\ref{sec:mu_deviation_demo} for more details.

As is visually apparent from Figure~\ref{fig:mu_gpe}, student grades in physics follow a different form compared to the other subject areas represented there.  
Compared to the non-physics courses, the two physics courses have intercepts at $T_0 = 0.9$ lower by $\sim 0.7$ grade points, the local slopes are nearly twice as large, and 
they have positive curvatures ($a_2 \sim +7$) whereas the other courses display negative curvatures ($a_2 \sim -3$). 
In essence, even students with the strong math abilities encounter a larger grade penalty in physics compared to these other subjects.

\section{Results}\label{sec:results}

Due to their similar structure, we analyze the combined sample of PHYSICS 140 and 240 student behavior here, beginning with deviations from mean expected GPE at a given math test score $T$ as a function of study volume. Because of the wide dynamic range and log-normal distribution of study volume (Figure~\ref{fig:hist_NQtot}), we employ $\log_{10}(\Nqtot)$ as the independent variable.  The use of a logarithmic measure of study enables a natural interpretation in terms of \emph{multiplicative factors} of student effort. 
We then investigate differences in study volume and GPE among demographic sub-groups. 
The section concludes with remarks on using GPAO instead of $T$ as a baseline condition for grade prediction.

\subsection{Evidence of grade gains from practice study} 
\label{sec:mu_deviation_study}

With the fit to mean GPE as a function of math test score $T$ in hand (see Figure~\ref{fig:mu_gpe}), we are now poised to address the question: ``For a given student, how do PR study habits relate to their final course grade?'' 
We focus on the shift in grade earned by a given student from expectations at a given $T$: 
\begin{equation}\label{eq:deltaGPE}
    \Delta_{{\rm GPE,}i} \ \equiv \ {\rm GPE}_i - \mu_{\rm GPE}(T_i). 
\end{equation}
We look to measure how these mean shifts in student grade depend on some measure of study volume $X$, meaning we seek $\langle \Delta_{\rm GPE} \rangle (X)$.


\begin{figure}[!ht]\centering
  \includegraphics [width=\linewidth] {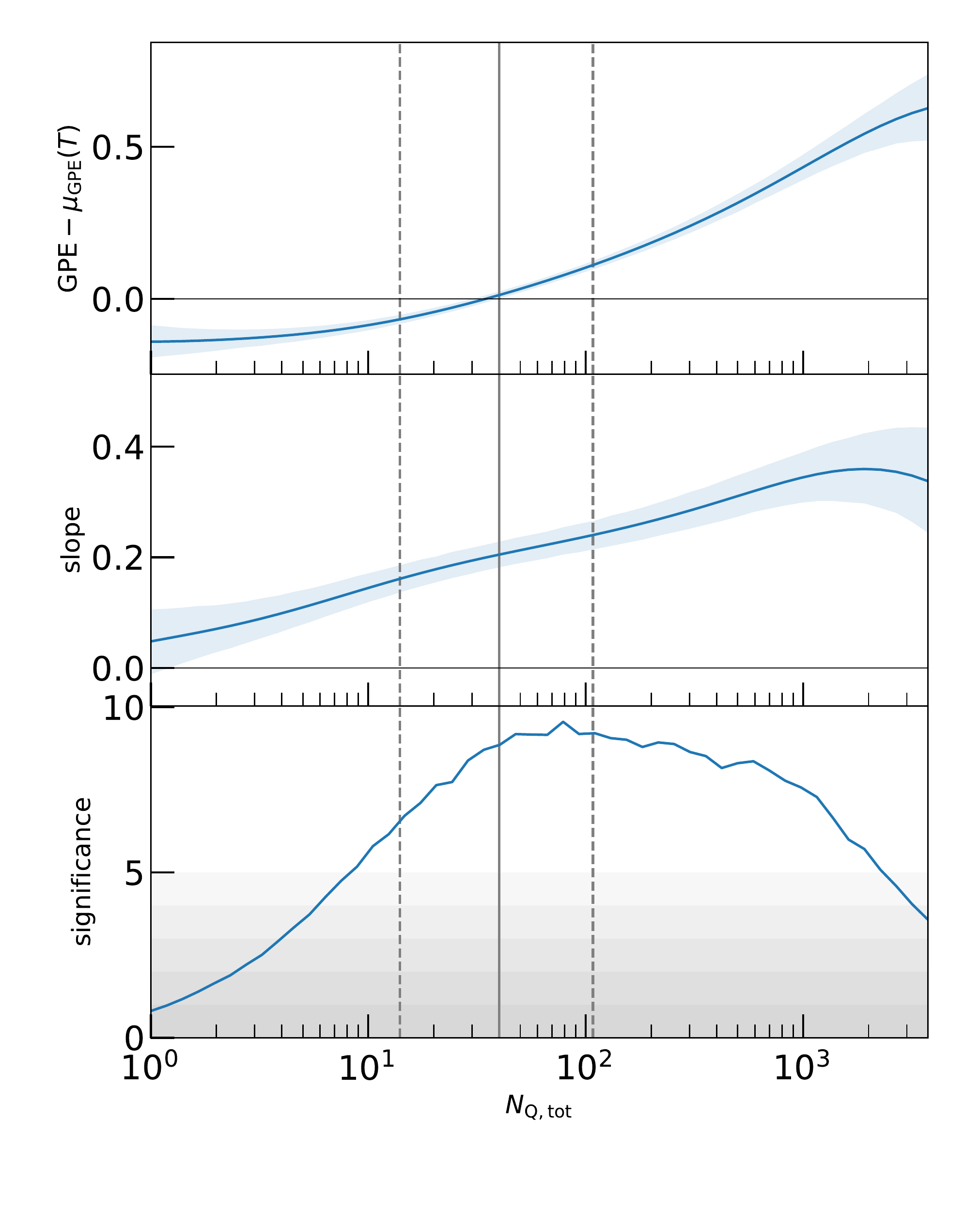}
  \caption{
    Grade gains as a function of practice effort for the combined sample of PHYSICS 140 and 240 students. Study volume is measured by the logarithm of the total number of PR questions encountered over the term, $\Nqtot$.  Vertical lines indicate quartile values of $N_{\rm Q, tot}$ for students who use PR; half practice between 14 and 108 problems, with a median of 40.  Statistics are computed with KLLR using a kernel width of $0.56$ in $\log_{10} \Nqtot$ (a factor of $3.6$) and shaded regions are $1\sigma$ uncertainties from bootstrap resampling.  
    {\bf Upper panel:} Mean shift in grade points earned, $\langle \Delta_{\rm GPE} \rangle$, relative to that expected from pre-college math score, equation~\eqref{eq:deltaGPE}. 
    {\bf Middle:} Local slope of $\langle \Delta_{\rm GPE} \rangle$. 
    {\bf Lower:} Significance level of a non-zero slope. 
    Note that students who didn't engage with PR are not included in this analysis.  
  }
  \label{fig:deviation_140240}
\end{figure}


Using KLLR, we measure mean final grade point deviation, equation~\eqref{eq:deltaGPE}, as a function of our primary study volume indicator, $X = \log_{10}(\Nqtot)$. KLLR kernel width, as calculated from equation~\eqref{eqn:KLLR_kw} is $\sigma_{\rm KLLR} = 0.56$ (a factor of 3.6 spread). 
We find a significant increase in mean grade points earned shown in the top panel of Figure~\ref{fig:deviation_140240}. 
The mean behavior monotonically grows from a value of $-0.2$ (a grade point penalty relative to the $T$-determined expectation) to $+0.6$ (a relative grade point benefit) as the total number of questions attempted over the term increases from one to over one thousand.  
The middle panel shows the local slope of the grade gain, which increases nearly linearly in $\log(\Nqtot)$, and 
the lower panel shows the significance of that slope being non-zero.  

The top panel shows the basic result that students who study more tend, on average, to do better in the course.  The novelty of this unsurprising finding lies in both the \emph{precise quantification} of mean grade gain as a function of term-aggregated study volume and its \emph{large overall amplitude}. 
These findings provide evidence-based answers to questions such as  
  ``How much should I study to get a half grade point boost in my final grade?'' (on average: approximately 660 questions over the term) or 
  ``If I study one question each school day, how much better am I likely do in the course than without study?'' (on average: roughly 0.24 grade points improvement). 
These outcomes hold true \emph{on average}; as discussed below, significant residual scatter in grade remains, reflecting a range of extraneous factors that influence student learning of physics.  

The KLLR characterization of study gains are largely in agreement with a simple quadratic trend, as seen in the near-linearity of the slope (middle panel). Fitting the trend in Figure~\ref{fig:deviation_140240} around the median value of $\Nqtot = 40$ we find mean behavior 
\begin{equation}\label{eq:DeltaGPEfit}
\langle \Delta_{\rm GPE} \rangle =  (0.00 \pm 0.01) + (0.22 \pm 0.02) \, \nu + (0.08 \pm 0.02) \, \nu^2,
\end{equation}
where $\nu \equiv \log_{10} \left( N_{\rm Q, tot} / 40 \right)$.  
\textsl{Grade gains from practice are thus quadratic in the log of number of questions attempted.}  While doubling study from one question a day to two a day will benefit a student, doubling from two a day to four will, on average, yield even more incremental benefit to their final grade. 

The lower panel of Figure~\ref{fig:deviation_140240} shows that the slope is $>3\sigma$ significantly positive for $N_{\rm Q, tot} \geq 4$.  The slope is roughly 0.1 at this point, but four problems over the course of a full term is a very minimal amount of practice, and grade earned at this level is indeed lower than the $T$-conditioned mean by 0.2 grade points. 
Gains in the top panel become more evident above ten questions, and the slope continues to grow, reaching 0.3 at $N_{\rm Q, tot} \simeq 300$.  This study volume represents roughly 25 questions per week, or five questions per school day. 
The positive slope in grade gain continues up to the limit of our study volume data.    
In our sample, 34 students, or $0.5\%$ of the population, attempted over 1000 questions within a semester, a volume that slightly oversamples the number of questions available and that corresponds to a rate of roughly 15 questions per school day.  

\begin{table}[!t]
\centering
  \caption{
    Maximum study gains, $\max(\Delta{\rm GPE})$, between high  and low PR study volume (see text) for each study measure along with the fractional reduction in grade variance after accounting for both $T$ and study volume. 
    The latter is calculated as $\Delta {\rm Var}_{0 \rightarrow i} \equiv ({\sigma_i}^2 - {\sigma_0}^2)/{\sigma_0}^2$, where $\sigma_0$ is the scatter in GPE originally and $\sigma_i$ is the scatter in grade point after removing trends in $\mu_{\rm GPE}(T)$ ($i=1$) and both that as well as $\langle \Delta_{\rm GPE} \rangle (X)$ for study measure $X$ ($i=2$), e.g. $X = \log_{10} \Nqtot$. 
    }
  \begin{ruledtabular}
    \begin{tabular}{llll}
      Indicator & $\max(\Delta_{\rm GPE})$ & $\Delta {\rm Var}_{0 \rightarrow 1}$ & $\Delta {\rm Var}_{0 \rightarrow 2}$ \\ 
      \hline
      \Nqtot & $+0.77 \pm 0.12$ & $-0.16 \pm 0.03$ & $-0.19 \pm 0.03$ \\
      \Nsess & $+0.61 \pm 0.19$ & $-0.16 \pm 0.03$ & $-0.18 \pm 0.03$ \\
      \Nqbar & $+0.72 \pm 0.11$ & $-0.16 \pm 0.03$ & $-0.19 \pm 0.03$ \\
    \end{tabular}
  \end{ruledtabular}
  \label{tab:deviation_trend_new}
\end{table}

Table~\ref{tab:deviation_trend_new} quantifies for each study volume indicator ($\Nqtot$, $\Nsess$, and $\Nqbar$) the overall grade gain, $\max(\Delta_{\rm GPE})$, defined as the maximum difference in KLLR mean values across the study volume domain.  Uncertainties in the difference are propagated from the respective KLLR bootstrap errors on the minimum and maximum values.  
The table also quantifies the fractional reduction in variance $\Delta{\rm Var}$ from the initial scatter in GPE $\sigma_0$ to the scatter after accounting for $\mu_{\rm GPE}(T)$ ($\Delta {\rm Var}_{0 \rightarrow 1}$) and after additionally accounting for $\langle \Delta_{\rm GPE} \rangle (X)$ in study measure $X$ ($\Delta {\rm Var}_{0 \rightarrow 2}$). The reduction in variance from the middle to end stage is then $\Delta {\rm Var}_{1 \rightarrow 2}$. 

For each study indicator, students with high study volume tend to do significantly ($> 3 \sigma$) better than those with low study volume. 
Across the three study indicators, we see similar gains (within $1\sigma$ of each other) between two-thirds and three-fourths of a full grade point. This result is consistent with moving from a B- to a B+ or from a B+ to an A. 
These results are also robust with varying kernel size: doubling or halving kernel width yields statistically identical results. 
Furthermore, the results are robust with varying student math ability. Dividing the student sample into terciles by $T$, we find similar overall grade gains: $0.60 \pm 0.21$, $0.73 \pm 0.19$, and $0.75 \pm 0.12$ for the lowest, intermediate, and highest $T$ terciles, respectively, averaged across study indicators. 
\textsl{All} students benefit similarly from increased study volume.

Across each study indicator, we see similar reductions of variance, with $\sim 20\%$ of the scatter in student grades explained by the joint $\mu_{\rm GPE}(T)$ and $\langle \Delta_{\rm GPE} \rangle (X)$ trends. 
The trend of mean grades running with $T$ accounts for the majority of this reduction, causing on its own a $16\% \pm 3\%$ reduction. 
The range of $\mu_{\rm GPE}(T)$ is roughly 1.2 grade points, so the running of $\langle \Delta_{\rm GPE} \rangle (X)$ with maximal running of roughly half the range has less capacity to reduce variance. 
Because the distribution of \Nqtot\ clusters strongly towards lower values, further limiting its ability to reduce variance. 
However, on sampling evenly in $\log \Nqtot$, we find a significant variance reduction of $\Delta{\rm Var}_{1 \rightarrow 2} = (9 \pm 3)\%$, with study explaining roughly 10\% of the remaining scatter, after accounting for $\mu_{\rm GPE}(T)$ expectations. 
This suggests that our initial inability to measure significant reduction of variance on accounting for study trends is tied to the clustering of \Nqtot\ at relatively modest values.

Below, we incorporate both math test score and study volume as fit here into a model of student grade and investigate divergences from model expectations for several demographic sub-populations.

\subsection{Demographic differences} \label{sec:mu_deviation_demo} 

\begin{figure}[!ht]\centering
  \includegraphics [width=\linewidth] {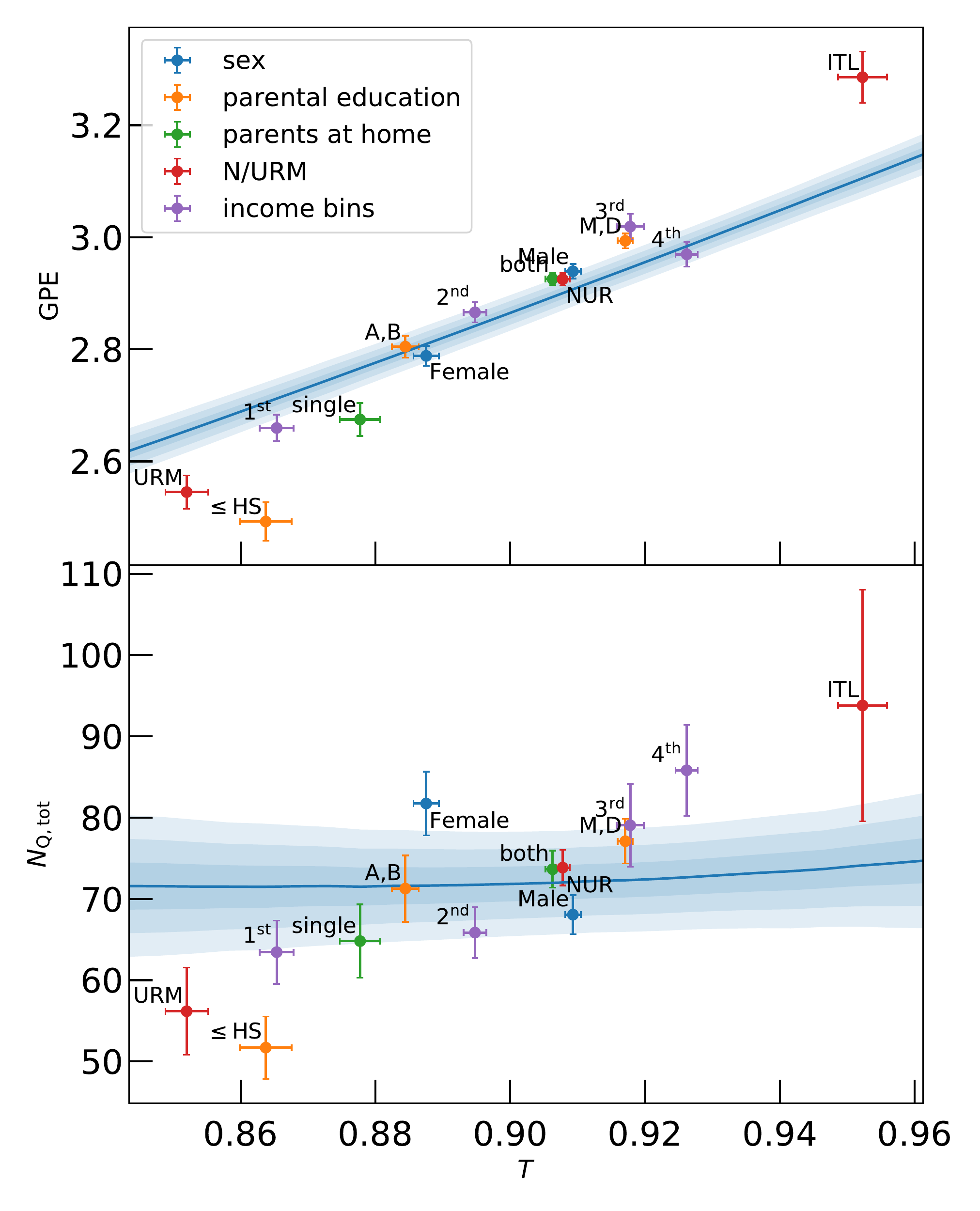} 
  \caption{
    Points show mean GPE, \Nqtot, and $T$ for the various demographic subpopulations indicated in the legend, with error bars showing standard deviation of the mean. 
    KLLR fits of GPE and $\Nqtot$ as functions of $T$ for the full student population, shown in blue with $\pm 1$, 2 and $3 \sigma$ uncertainties, demonstrate starkly different trends.  For the student population as a whole, grades strongly correlate with $T$ while study volume does not. 
  }
  \label{fig:mu_gpe_demo} 
\end{figure} 

On average, differing educational experiences exist across socioeconomic status, sex, race, ethnicity, home environment, and other factors.\cite{Pascarella+04,Delaney+11,McLanahan+13,Stroub+13,Owens+16} 
As detailed in \S\ref{sec:demo_groups}, we consider several of these demographic indicators (see Tables~\ref{tab:demographics} \&~\ref{tab:demo_pcts}) to the extent and precision available to us; most indicators are self-reported. 
We investigate the degree to which demographic groups are stratified in the space of $T$, GPE, and $\Nqtot$, then quantify how each sub-group's mean grade deviates from expectations.

As mentioned in \S\ref{sec:mu_fit}, we first investigated whether $\mu_{\rm GPE}(T)$ had any significant differences in form between demographic sub-groups. While there were significant differences in vertical intercept between sub-groups, there were no significant differences in slope nor curvature in the quadratic fitting of the courses. Compared to the global parameterization of Table~\ref{tab:mu_fits_new}, sub-population fits for curvature were consistent to $<2\sigma$ and slopes were consistent to $<3\sigma$. In contrast, there were several significant differences in offset $a_0$ for three groups, with the $\leq$HS group falling $\sim 0.3$ points below expectations while URM students and students of single-parent households both fell $\sim 0.1$ points below expectations. 
Once more, we emphasize that these shifts are precisely what this study seeks to quantify, asking what correlates with vertical shifts from mean expected grade at a given $T$. Now that we have shown that the fitting of $\mu_{\rm GPE}(T)$ is robust to demographic sub-group, we turn our attention to how demographic sub-group deviation from expectations relate to study volume. 


Figure~\ref{fig:mu_gpe_demo} illustrates the relationship between mean measures for demographic groups, shown as points, and the overall trend for all students, shown as the line (with shaded regions indicating $\pm 3 \sigma$ uncertainty on the KLLR fit).  
For this analysis, we measure mean $\Nqtot$ values for all students for whom we have $T$ scores, including those for whom $\Nqtot=0$.  Note that the mean value, $\langle \Nqtot \rangle \simeq 72$, differs from the median log value of 40 shown in Fig.~\ref{fig:deviation_140240} because the distribution of $\Nqtot$ is close to log-normal and has substantial width.  This level of effort corresponds to attempting roughly five questions per week during the term.  

GPE differs starkly from \Nqtot\ in its trend with $T$. 
Mean grade earned (top panel) correlates strongly with math test score ($0.38 \pm 0.01$), rising by half a grade point over just a 0.1 increase in $T$.  
In stark contrast, the volume of study by students (lower panel) is effectively flat (insignificantly correlated: $0.02 \pm 0.01$); low $T$ students study very nearly as much as their high $T$ counterparts.  We return to this issue when considering GPAO as an alternate to $T$ below.

Mean values of math test score $T$ for demographic groups cover the $0.1$ domain shown, with students whose parents have the lowest education level ($\le$HS) and underrepresented minority (URM) students possessing the lowest scores and international students (ITL) the highest. These groups deviate somewhat from the overall trend in grade, with ITL students lying above and URM \& $\le$HS below the population mean. Note that, because of the positive curvature in mean grade with $T$, mean values of sub-populations will tend to lie somewhat above the trend line.  

While students across the spectrum of math test scores put in similar level of practice effort, some differences between demographic groups are apparent in the lower panel of Figure~\ref{fig:mu_gpe_demo}. Though deviations are generally less significant than in the upper panel, URM \& $\le$HS populations tend to rest below the trend whereas females, high income groups, and international students rest somewhat above. 




\subsubsection*{Does Math Score and Study Volume Explain Demographic Grade Shifts?} \label{sec:mu_model_demo}

The trends in the bottom panel of Figure~\ref{fig:mu_gpe_demo} suggest an explanation for demographic grade deviations; vis. that the demographic shifts in study volume often have similar sign to those of $\Delta_{\rm GPE}$. 
However, these shifts sometimes misalign; for example, the highest income group tends to study more than average yet achieves only average grades. 
To address the degree to which study explains demographic deviations from mean trends, we employ a model that combines the expected mean grade conditioned on $T$ combined with the grade shift as a function of study volume using the full student population (KLLR fits of Figures~\ref{fig:mu_gpe} \&~\ref{fig:deviation_140240}, largely equivalent to the parameterized fits of equations~\eqref{eqn:mu_gpe} \&~\eqref{eq:DeltaGPEfit}). 

%
%
%
%

\begin{figure}[!ht]\centering
  \includegraphics [width=\linewidth] {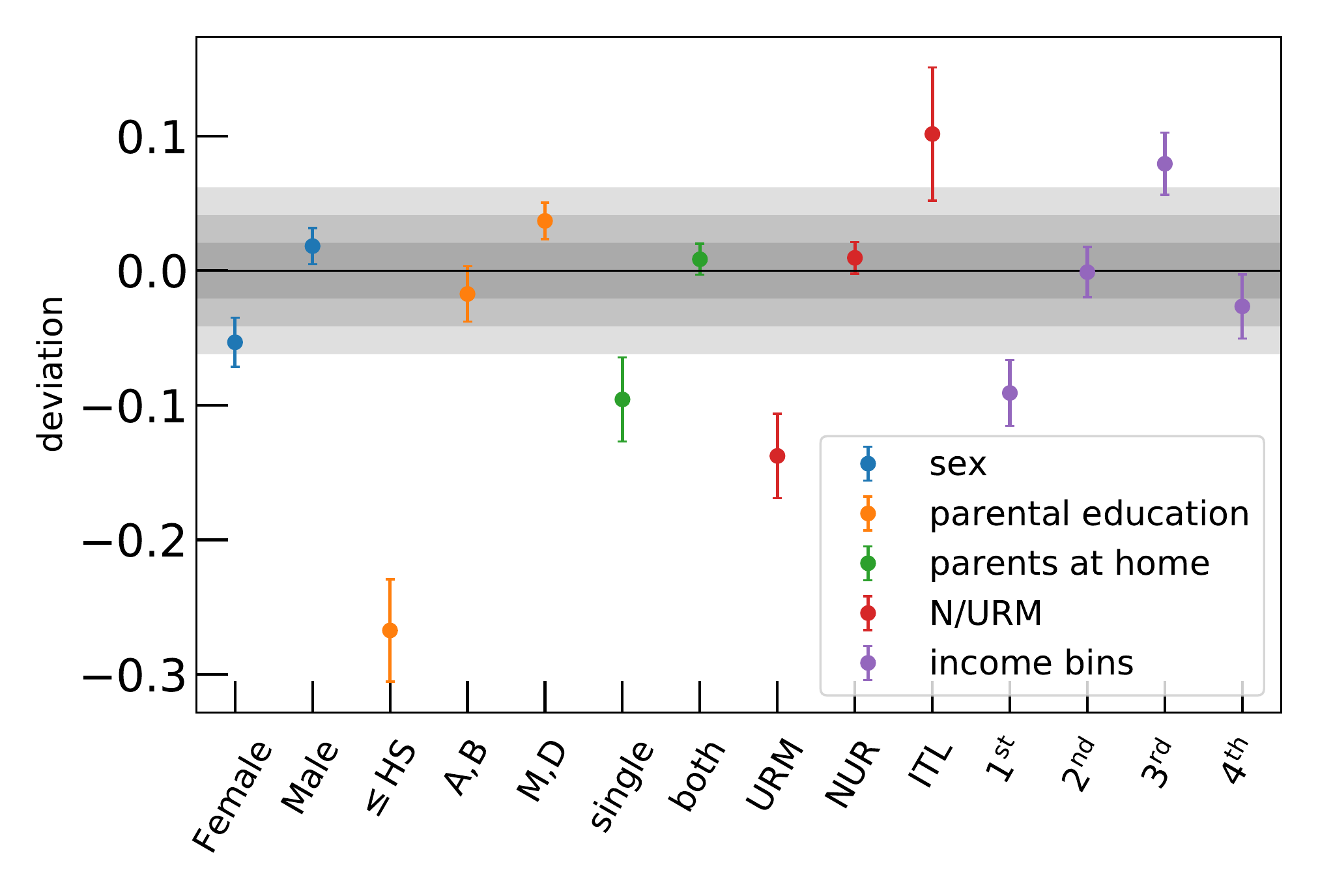}
  \caption{
    Mean GPE deviations of demographic groups from the overall trend based on combining math score $T$ and study volume $\Nqtot$ for the full student population.   
    The shaded region around the null line shows $\pm 3 \sigma$ uncertainties of the predictions based on combining $T$ and $\Nqtot$. 
    The error bar for each demographic group is the $\pm 1 \sigma$ error in that group's mean GPE deviation from the mean model prediction. 
  }
  \label{fig:cf_diffs}
\end{figure}

Figure~\ref{fig:cf_diffs} shows the mean deviation from these expectations for each demographic sub-population. The zero line and associated median $\pm 3\sigma$ uncertainties  (median combined uncertainties of the KLLR fits to $\mu_{\rm GPE}(T)$ and $\langle \Delta_{\rm GPE} \rangle (\log_{10} \Nqtot)$ for a median student) represent the expected value based on our model of the full student population. 
Most demographic sub-populations have mean deviations consistent with the model's expectations, but significant outliers are seen toward lower math scores. 
%
Students whose parents had a high school degree or less lie significantly below expectations, at $-0.27 \pm 0.04$ grade points, 
as do URM students, at $-0.14 \pm 0.04$ grade points. 
In addition, students from households with only a single parent present as well as students from high schools in the lowest income regions display a smaller, less significant deficit of $-0.09 \pm 0.04$ grade points. 
Though the third income bin is slightly above expectations ($+0.08 \pm 0.03$), the fourth income bin is slightly below. 
All other sub-populations were $<2\sigma$ deviant from expectations.

\subsection{Comparison to GPAO} \label{sec:GPAO}
In this section, we investigate results when using GPAO as a baseline for comparison, rather than $T$. First, we explain some limitations of GPAO (with more issues outlined in Appendix~\ref{apx:GPAO}), Second, for PHYSICS 140 and 240 we show the $\mu_{\rm GPE}({\rm GPAO})$ plot analogous to Figure~\ref{fig:mu_gpe}. Third, we show for the two courses combined a gains of study table, analogous to Table~\ref{tab:deviation_trend_new}.

\subsubsection{GPAO correlation with study} \label{sec:corr}
A core issue with using GPAO as a baseline for comparison (as is done, for example, with the ``grade anomaly'' ${\rm GPE} - {\rm GPAO}$ of \citet{Weaverdyck+20}) is its correlation with study habits. Averaged over log or non-logged versions of all three study volume indicators used in this study (a total of six possible metrics), correlations between study and GPAO are statistically significant, $0.132 \pm 0.013$, whereas correlations between study and $T$ are insignificant, only $0.013 \pm 0.013$. 
Regardless of study metric, GPAO consistently shows stronger and more significant correlations to study than $T$ does. Because we seek to measure how study volume influences final grade, this coupling makes GPAO relatively untenable compared to $T$ for this purpose.

\begin{figure}[!t]\centering
  \includegraphics [width=\linewidth] {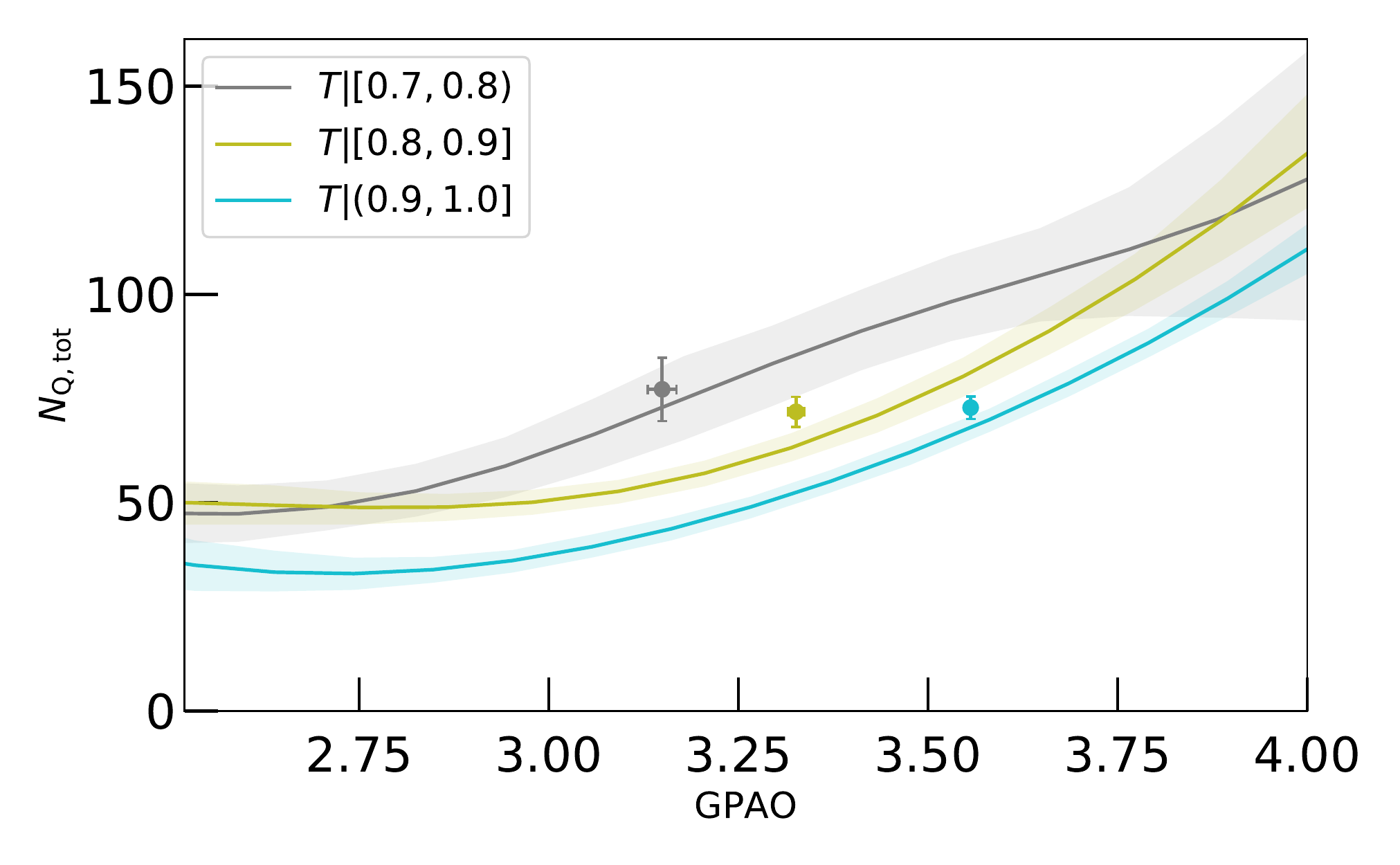}
  \caption{
    Study volume as a function of GPAO for students in bins of $T$ shown in the legend. 
    Mean values are KLLR-derived using a kernel width $0.40$ 
    and $1\sigma$ uncertainties are shown as shaded regions. 
    Population means for each $T$ group are shown as points with $\pm1\sigma$ error bars. Significant study trends with GPAO exist within each math score range. 
    %
  }
  \label{fig:N_sess__vs__GPAO__by__T}
\end{figure}

Figure~\ref{fig:N_sess__vs__GPAO__by__T} shows the interplay between PR study volume, 
GPAO, and composite test score $T$ for PHYSICS 140 and 240 combined. 
KLLR fits to study volume as a function of GPAO show clear rising behavior for each individual $T$ bin, while the means (displayed as points with error bars) show no significant trend in study behavior with $T$. 
The striation of trend lines shows that among students with the same GPAO, those with lower $T$ scores tended to study more on PR. 
In contrast to the lack of correlation between $T$ and study, at \emph{fixed} GPAO there is a significant trend of decreasing PR study volume for students with higher math scores, $T$. 

Though students with high $T$ scores tend to have higher grades and though students who studied more tended to have higher grades, there was no significant correlation between $T$ and study volume. This seeming inconsistency is resolved in the anticorrelation between study volume and $T$ \emph{at fixed GPAO}. 
If we take PR study volume as a proxy for typical student study habits, then this could perhaps be interpreted to reveal that students with higher $T$ scores tended to not need to study as much to receive the same high grades as students with lower $T$ scores.

Because study volume correlates to GPAO, its use would at least partially undermine our attempt to measure the benefits of study. Due to its correlation, any study gain measured using GPAO as baseline instead of $T$ as baseline would show reduced study gains at fixed GPAO. We proceed now to quantify this reduction in measured gains.


\subsubsection{Fitting GPE as a function of GPAO}
We begin by fitting student grades earned in each physics class as a function of their end-of-term grade point average in all other courses (GPAO). As in \S\ref{sec:mu_fit}, we use a KLLR fitting to discern the general trend from a relatively agnostic viewpoint. 

\begin{figure}[!t]\centering
  \includegraphics [width=\linewidth] {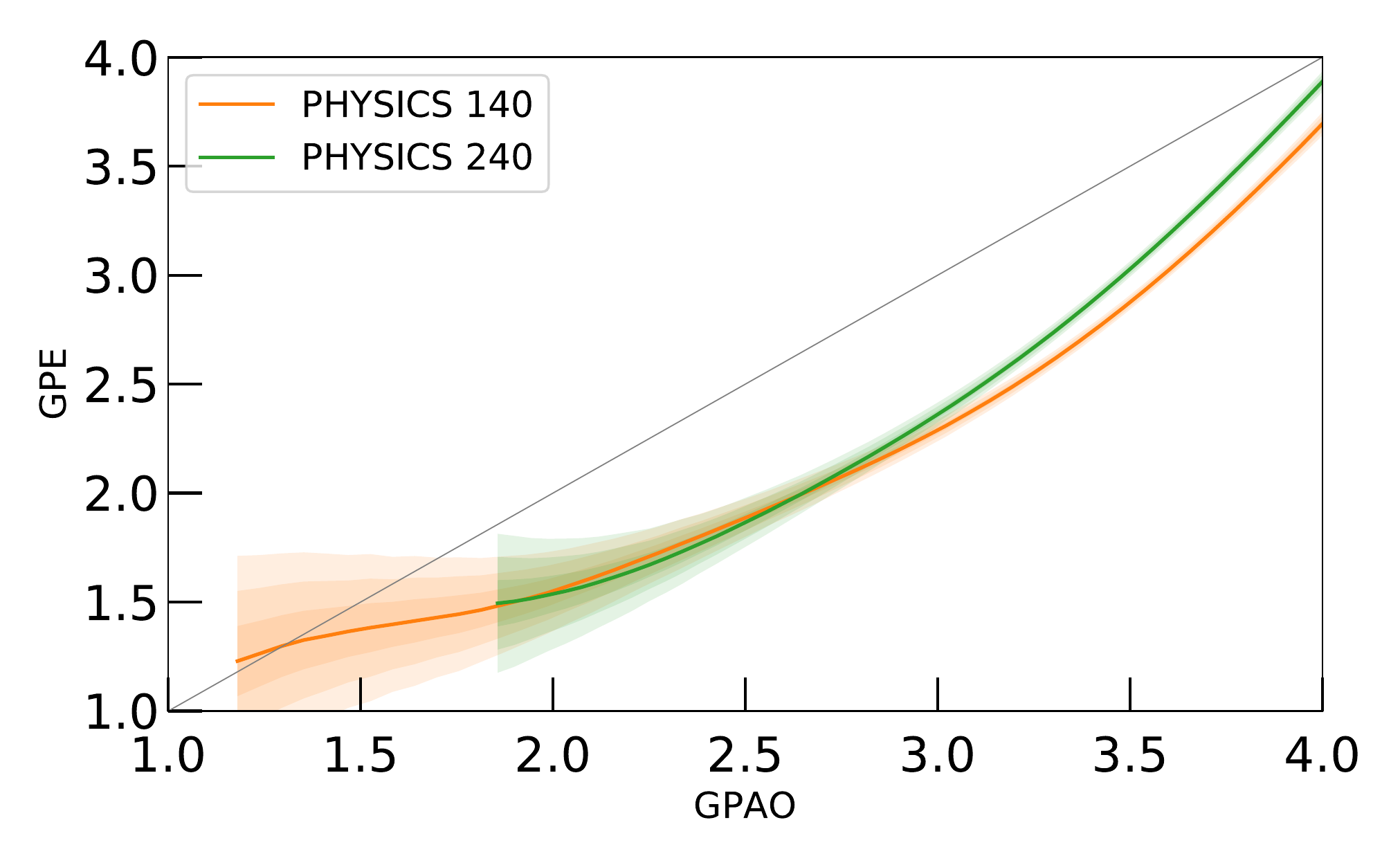}
  \caption{
    As Figure~\ref{fig:mu_gpe}, but fitting mean GPE for the two physics courses as a function of GPAO rather than $T$. Grey line indicates equality, where a student's grade earned in physics would equal their average grades earned from other courses. 
    The GPAO distribution was similar between courses, with means around $3.4$, standard deviation of $\pm 0.5$, and ${\rm GPAO} \gtrsim 2.45$ for 95\% of students. 
  }
  \label{fig:mu(GPAO)}
\end{figure}


Figure~\ref{fig:mu(GPAO)} shows the KLLR fits of mean GPE as a function of GPAO for both physics courses individually. 
As with $\mu_{\rm GPE}(T)$, we see significant positive curvature. The bulk of the population has ${\rm GPE} < {\rm GPAO}$ (diagonal line), meaning that these physics courses tend to be the lower grades on their transcripts, especially so for students with low GPAO already. For example, a student with otherwise straight As (${\rm GPAO} = 4.0$) would likely get an A- while a student with otherwise straight Bs (${\rm GPAO} = 3.0$) would likely get a C+.

\subsubsection{Deviations from GPAO mean due to study volume}
After calculating a deviation from expected grade at a given GPAO, $\Delta = {\rm GPE} - \mu_{\rm GPE}({\rm GPAO})$, we investigate whether a trend exists in this deviation as a function of PR study volume, using KLLR fitting, analogous to that done in Figure~\ref{fig:deviation_140240}. We present gains of study in Table~\ref{tab:deviation_trend_GPAO}, analogous to Table~\ref{tab:deviation_trend_new}. 
Those with highest study volume tended to do better than expected by $0.28 \pm 0.07$ grade points (averaging over the three study indicators) 
as compared to those with low study volume.

\begin{table}[!ht]
\centering
  \caption{
    As Table~\ref{tab:deviation_trend_new}, but measuring gains and variance reduction with GPAO as a baseline for grade expectations (rather than $T$). 
    Note that variance decreases from $1 \rightarrow 2$ are all quite insignificant. 
  }
  \begin{ruledtabular}
    \begin{tabular}{llll}
      Indicator & $\max(\Delta_{\rm GPE})$ & $\Delta {\rm Var}_{0 \rightarrow 1}$ & $\Delta {\rm Var}_{0 \rightarrow 2}$ \\ 
      \hline
      \Nqtot & $+0.302 \pm 0.080$ & $-0.467 \pm 0.010$ & $-0.469 \pm 0.009$ \\
      \Nsess & $+0.207 \pm 0.059$ & $-0.467 \pm 0.010$ & $-0.468 \pm 0.010$ \\
      \Nqbar & $+0.335 \pm 0.053$ & $-0.467 \pm 0.010$ & $-0.471 \pm 0.009$ \\
    \end{tabular}
  \end{ruledtabular}
  \label{tab:deviation_trend_GPAO}
\end{table}

This is less than half the grade gain measured in Table~\ref{tab:deviation_trend_new}, where $T$ was used as a baseline for comparison. The stark difference in grade gains measured suggests that the significant correlation between GPAO and study volume (see \S\ref{sec:corr}) has subsumed a large portion of the grade gains measured. That is, because GPAO reflects in part an individual's study volume, subtracting it out in the initial baseline grade estimate $\mu_{\rm GPE}({\rm GPAO})$ removes a significant fraction of the measurable grade gains due to study. Despite GPAO predicting grades better than $T$ (reducing variance by 47\% rather than only 19\%), GPAO has less utility in quantifying the grade benefits of study, washing out over half of the measurable grade benefit.



\section{Discussion}\label{sec:discussion}




Our findings help demonstrate to educators and learners the benefits of practice study in introductory physics courses. While this general finding is neither original nor surprising, its precise quantification is novel, and the overall magnitude in grade gain for students classified by pre-college math ability is large.  

The low-stakes formative assessment of PR matches well to the summative midterm and final examinations of the two physics courses considered here. After all, these PR courses consist of nearly 1000 problems used on past examinations in each course. 
Given the median response time per problem of roughly 90 seconds\cite{Weaverdyck+20}, we do not think that memorization \textsl{per se} plays an important role. As mentioned earlier, exams consist of entirely novel questions (though certainly with some structural and topical overlap), so literal repetition of PR questions is quite unlikely. 
Rather than memorization, other factors such as the similarity of problem construction and structure, the mix of quantitative and qualitative questions, and particular ``tricky'' problem styles involving multiple physics concepts are likely to be similar between old and new exams.



Our findings with respect to demographic characteristics motivate more study of how better to support physics learning for first-generation (our $\le$HS category) and underrepresented minority students (our URM category). 
Students from single-parent and low-income households also earn lower grades than expected (Fig.~\ref{fig:mu_gpe_demo}). 
These categories overlap, inviting future work to understand how students with multiple of these identities perform.\cite{Saw+18}


Our comparison to GPAO study trends reveals far more grade benefit measured when conditioning on $T$ than on GPAO. Because GPAO significantly correlates with study volume (while $T$ does not), subtracting out a GPAO baseline from grades 
subtracts out the effects of study behaviors to some extent.
While GPAO may be the most accurate predictor of GPE, it is a flawed baseline for the purposes of our analysis (see also Appendix~\ref{apx:GPAO})  
as it washes away the majority the the study benefit trend measured in \S\ref{sec:mu_deviation_study}. 
We thus find utility in using $T$ as a baseline for student comparison, rather than GPAO.

The utility of $T$ could in part be due to its significant correlation with general mental ability $g$.\cite{Frey+04,Koenig+08,Coyle+08,Sackett+08} 
As $g$ is largely intrinsic (polygenic, yet $\gtrsim 50\%$ heritable and $\gtrsim 90\%$ consistent with age\cite{Jensen98,Neisser+96,Bouchard_2013,Panizzon+14}) to each student, it is largely orthogonal to a student's external, personal choice of study habits (dependent more so on personality than on $T$; see \S\ref{sec:corr}). 
In contrast, GPAO represents a mix of both internal ability and external choice, and thus subsumes a portion of the study signal we attempt to measure in this study.

Another potential reason for the strong utility of $T$ could lie in the similarity of assessment styles. Both standardized tests and the assessment methods used in many large introductory STEM courses take the form of high-stakes multiple-choice questions in a timed setting.

Regardless of interpretation, this difference in grade gains (measured with GPAO vs. $T$ baselines) warrants further investigation and discussion. As more colleges move to ``test-optional'' admissions in the future, we may in the process be limiting the potential of future investigations to quantify student learning gains.

\section{Concluding Summary}\label{sec:summary}

Using a large sample of practice study from the Problem Roulette service, we find that student final grades scale quadratically with the logarithm of the term-aggregated number of questions encountered, with an overall gain of nearly 0.8 grade points. 
By comparison, the largest demographic deviation we find is $\lesssim 0.3$ grade points, so with roughly a question per day on average, this deficit can be overcome purely through study.  

We summarize our findings in the following advice to students and teachers. 
Our advice to students is: 
\begin{quote}\emph{
  Do a problem for every time you brush your teeth. 
  Attempting at least one or two problems each school day of the term will likely lift your grade by one-quarter point. 
  Quadratic gains with log study implies that while doubling study volume from one to two questions per day benefits you, doubling again from two to four questions per day benefits your grade even more!
}\end{quote}

\noindent Our advice to teachers is: 
\begin{quote}\emph{
  Encouraging students to do at least one problem a day puts them above the median study habit; on average, this will land them above their expected grade. 
  Study can benefit students by up to three-quarters of a letter grade. 
  Differences between demographic groups, after accounting for $T$ and study, were $\lesssim 0.3$ grade points; first-generation students were the most significant outlier. 
  %
}\end{quote}

This study only scratches the surface of PR data available. Besides only working with physics courses, our model was relatively simple, treating $T$ and ${\rm GPE}$ separately and using other demographic, academic, and study-related variables only tangentially. A more nuanced model such as multi-level modeling or simultaneous fitting of many variables at once (such as with the machine learning tool SHAP\cite{Lundberg+17}) could help disentangle which factors are more or less causal (though it could not determine absolute causality). 
Future analyses should also treat nationality, ethnicity, and race more carefully, rather than using the broad categories of nationality and binary URM status. 
We also would like to investigate the effects of study session length---early results suggest that working for more than about 50~minutes tends to yield minimal gains. 
Finally, we wish to compare findings between all courses, observing which trends persist cross-subject.

We should never lose sight that the purpose of this work is to help individual students. Teachers and students alike can benefit from understanding how $T$ and study affect their grades. Teaching is a handshake, requiring earnest participation of both parties for best results. (We can't fall into the trap of subscribing to either a student deficit model or a teacher deficit model alone; both parties need to improve their habits and grow as individuals.) As students improve their scholastic habits and teachers improve course structure and learn how to best reach struggling individuals, we can grow together and improve the education system.

\appendix 

\section{Problem Roulette structure} \label{apx:PR}
After logging in and selecting a course, students can choose from various modes of study: Individual, Group, or Practice Exam (student-generated or faculty-generated as formal practice exams). 
Problem Roulette also collects questions from topics that students struggle with most, which students can study from. 

In the Individual or Group study modes, students select which topics to study (e.g., ``Vector Algebra'', ``Relative Motion''), whether or not to use a timer for the session, and how many questions to pull (defaults: 10, 25, all). They are then presented with an exam-like set of questions one at a time. After the session, they are presented with a review of the questions attempted with correct answers indicated, along with an overall accuracy score and the session duration. 

Instructors can view course analytics for a given term. At the instructor dashboard, they see student activity (e.g., which days saw more PR use), student accuracy versus questions answered for each topic, the accuracy of answers by topic, and the accuracy of individual questions answered. This information can reveal which topics or types of questions students tend to struggle with most or least.

\section{Additional issues with GPAO} \label{apx:GPAO}
Though GPAO correlates more with course grade earned (GPE) than any other indicator we investigated, it has a several issues. 
The most crucial issue of its correlation with study is detailed in \S\ref{sec:GPAO}; here we delineate several additional issues with using GPAO as a baseline for GPE comparison (vs. $T$ as a baseline, as used in this paper), that is, using grade anomaly ${\rm GPE} - {\rm GPAO}$ instead of the metric ${\rm GPE} - \mu_{\rm GPE}(T)$ used herein.

\subsection{Relation to personality traits}

Personality traits correlate significantly with grades,\cite{koseoglu2016} mediated through study habits.\cite{Aluja_Blanch_2004} 
Because GPAO depends on study and some demographic populations have different mean study behaviors than others, significant differences exist between some different demographic sub-populations. 
For example, male students tend to study less than female students, and they tend to have a lower GPAO than female students by roughly 0.07 to 0.14 points, despite having significantly higher $T$ scores on average.\cite{Richardson+12} This means that in a class where each student got identical grades, the grade anomaly for males and females could still differ for causes \emph{extrinsic} to the course in question.

\subsection{Differing course selection}
GPAO depends on course selection of individual students, but course selection differs drastically by demographics, and not all course loads are equal. If all students took similar courses regardless of demographic group, then GPAO would be a fair measure of ``anomaly''---how differently the course was graded compared to expectations from other courses. However, different groups tend to take different courses, leading to systemic shifts between different demographic sub-groups. If a student tends to take easier courses, they could have a 4.0 GPAO, then this would allow the student more free time to study on PR. However, a similar signal comes from a student who, despite taking incredibly challenging courses, works very hard and still has a 4.0 GPAO, yet then has very little free time to study on PR. While this exemplifies why GPAO has power in predicting GPE, it also shows that interpretation of GPAO is clouded. 


\subsection{Deletion of campus-wide trends}
Grade anomaly ${\rm GPE} - {\rm GPAO}$ wipes out campus-wide biases. Expressing this numerically: if every course in the school consistently awarded lower grades by a bias $b$ to a given group $g$ versus the rest of the population $p$, then the grade anomaly of that group would be 
\begin{align}
  (\text{grade anomaly})_g 
  &= {\rm GPE}_g - {\rm GPAO}_g \\
  &= ({\rm GPE}_p - b) - ({\rm GPAO}_p - b) \nonumber \\
  &= {\rm GPE}_p - {\rm GPAO}_p 
  = (\text{grade anomaly})_p, \nonumber
\end{align}
so the group then has identical grade anomaly to the rest of the population. This then erases the effects of campus-wide biases.

\begin{acknowledgments}

WKB was partially funded by an Academic Innovation Fund award from the UM Center for Academic Innovation (CAI).
The data were obtained from CAI and the UM Learning Analytics Data Architecture (LARC) database. 

\end{acknowledgments}




\section*{Bibliography}

\bibliography{main} 
\end{document}